\documentclass[aps,prb,twocolumn,epsfig,floatfix,showpacs,superscriptaddress,amsmath,amssymb]{revtex4}
\usepackage[T1]{fontenc}
\usepackage[latin9]{inputenc}
\usepackage{amsmath}
\usepackage{graphicx}
\usepackage{amssymb}
\usepackage{amsfonts}
\usepackage{mathrsfs}

\begin{document}

\title{Topological Peierls Transitions in M\"{o}bius Molecular Devices}

\author{Z. R. Gong}

\affiliation{Institute of Theoretical Physics, The Chinese Academy of Sciences,
Beijing, 100080, China}

\author{Z. Song}

\affiliation{School of Physics, Nankai University, Tianjin, 300071,
China}

\author{C. P. Sun}

\affiliation{Institute of Theoretical Physics, The Chinese Academy
of Sciences, Beijing, 100080, China}

\date{\today }
\begin{abstract}
We study the topological properties of Peierls transitions in a
monovalent M\"{o}bius ladder. Along the transverse and longitudinal
directions of the ladder, there exist plenty Peierls phases
corresponding to various dimerization patterns. Resulted from a
special modulation, namely, staggered modulation along the
longitudinal direction, the ladder system in the insulator phase
behaves as a {}``topological insulator'', which possesses charged
solitons as the gapless edge states existing in the gap. Such
solitary states promise the dispersionless propagation along the
longitudinal direction of the ladder system. Intrinsically, these
non-trivial edges states originates from the Peierls phases
boundary, which arises from the non-trivial $\mathbb{Z}^{2}$
topological configuration.
\end{abstract}


\maketitle

\section{\label{sec:one}INTRODUCTION}

Many recent efforts have been made to both experimental and theoretical
investigations of the application oriented molecular devices~\cite{MD1}.
As new type of the quantum coherence devices, molecular device emerges
various novel quantum effects, which enlarge the ranges of the material
design~\cite{device1,device2,device3,device4}. In these systems,
the exotic quantum features would be induced by the non-trivial topology,
and the observable quantum effects can also be used to specify the
topological constructions of the system. Indeed, this non-trivial
topology induced quantum effects never appear in the topologically
trivial systems.

With non-trivial topology, the twisted boundary condition in the
M\"{o}bius strip is a good subject to demonstrate the significant
role of topological structure in low-dimensional
physics~\cite{MBC1,MBC2}. The M\"{o}bius strip is a non-orientable
manifold, whose edge defines a two-point bundles over
$\mathbf{S^{1}}$ and thus $\mathbb{Z}^{2}$ topological
configuration. This simple but topologically non-trivial system
possesses mathematically rigorous description and the
accessibilities of the experimental realization. Actually, the
M\"{o}bius boundary condition has been synthesized in the aromatic
annulenes, nanographite ribbons and conjugated
polymers~\cite{MS1,MS2,MS3,MS4,MS5}. These progresses motivate us to
propose a tight-binding quantum device with M\"{o}bius topology and
investigate its quantum properties of the transportation of the
spinless particles~\cite{Zhao,Guo}.

Another important phenomenon in molecular devices~\cite{phonon1,phonon2}
is their Peierls instability, which exists universally in low-dimensional
physical system including polymers, spin chains, and organic materials,
etc~\cite{PT1,PT2}. The significance of the Peierls transition is
that after the lattice is deformed due to the electron-phonon coupling,
the system energy is decreased and thus the changed energy band structure
converts the original metal phase into an insulator one. Actually,
the existence of metal-insulator transition in the polyacetylene~\cite{PT1,PT2}
originates from this lattice modulation.

For the M\"{o}bius ladder configuration as a quasi-one dimensional
(Q1D) system, investigation of the Peierls transitions apparently
combines both the topological effect and the structure instability.
In this paper, we demonstrate the various dimerization patterns in
the monovalent M\"{o}bius ladder in details. Because of the
possibilities of the lattice deformations along transverse and
longitudinal directions, there exist five typical uniform
dimerization patterns that we will display in this paper. In
contrast, it is noticed that there is only one dimerization pattern
in a one dimensional system. All the five dimerization patterns
contain the rung, the columnar, the staggered dimerization patterns
and the vertically saw-toothed, the inclined saw-toothed
dimerization patterns as the combinations of the former three ones,
all of which will be explicitly defined in the next section.

We compare the Peierls phase diagrams of the M\"{o}bius ladder with
that of the generic one. Here, the generic ladder satisfies periodic
boundary condition. It is discovered that when the generic boundary
condition was replaced by a M\"{o}bius one, the conducting
properties are dramatically changed for the staggered and the
inclined saw-tooth dimerization patterns and not changed at all for
the other three patterns. This fact motivates us to use the
continuum model to analyze the exotic dimerizations. We notice the
existence of the localized state~\cite{ES1,ES2}, and find the
charged solitons propagating in the bulk, which promise that the
M\"{o}bius ladder with staggered dimerization is eventually
metallic. Being similar to the gapless localized states in graphene
strip, we also point out that our model behaves like a
{}{}``topological insulator\textquotedblright{}~\cite{TI1,TI2} with
localized state existing at topological non-trivial boundary. Here,
the topological insulator refers to a bulk insulator which possesses
robust metallic localized states, which is different from mundane
band insulator. These localized states are actually $\mathbb{Z}^{2}$
topologically invariant, which characterize the time-reversal
invariance of the topological insulator.

This paper is organized as following. In Sec.~\ref{sec:two}, we
present the lattice Hamiltonian of the M\"{o}bius ladder and
calculate the Peierls phase diagram. In comparison with the
M\"{o}bius case, we also calculate the Peierls phase diagram of the
generic ladder. To prove the existence of the non-trivial localized
states, we introduce the continuum model of the M\"{o}bius ladder in
Sec.~\ref{sec:three} as well as the one of the generic ladder
without any solitonary solution. We conclude our main results in
Sec.~\ref{sec:four}. The detailed derivation of the continuum model
from the lattice model is shown in Appendix \ref{app:one}.

\section{\label{sec:two}TOPOLOGICAL PEIERLS TRANSITIONS AND CORRESPONDING
PHASE DIAGRAM}

\subsection{\label{subsec:2-1}Model setup and dimerization patterns}

%
\begin{figure}[ptb]

\begin{centering}
\includegraphics[bb=16 366 589 767,clip,width=3.5in]{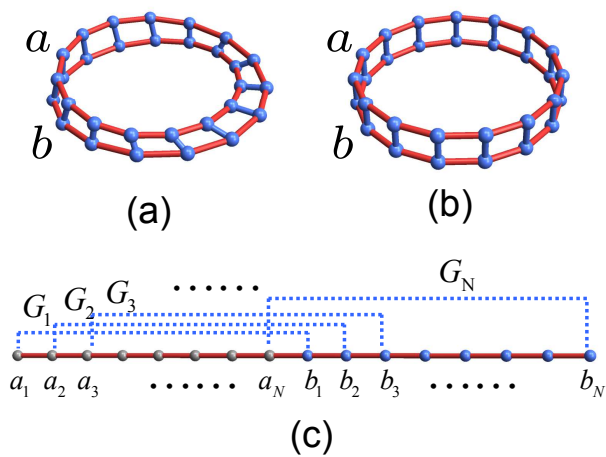}
\par\end{centering}

\caption{(Color online) Schematic illustrations of the ladder with
(a) M\"{o}bius boundary condition, (b) generic boundary condition
and (c) the corresponding one-dimensional version of M\"{o}bius
ladder system with long range coupling.}

\label{fig:fig1}
\end{figure}


In this section, we describe a tight binding model for the electrons
hopping on a ladder with the tight-binding Hamiltonian \begin{equation}
H_{\mathrm{e}}=\sum_{j=0}^{N-1}\mathbf{A}_{j}^{\dag}\mathbf{M}_{j}\mathbf{A}_{j}-\sum_{j=0}^{N-1}\mathbf{J}_{j}\mathbf{A}_{j}^{\dag}\mathbf{A}_{j+1}+h.c.,\label{2-1}\end{equation}
where operator-value vectors $\mathbf{A}_{j}=(a_{j},b_{j})^{T}$ is
defined in terms of the annihilation operator $a_{j}$ and $b_{j}$
of the upper and the lower chain of the ladder (see Fig.~\ref{fig:fig1}(a)),
which are denoted as $a$-chain and $b$-chain in the following discussion.
The transition matrices \begin{equation}
\mathbf{M}_{j}=\varepsilon_{j}\sigma_{z}-G_{j}\sigma_{x}\label{2-1-1}\end{equation}
 and $\mathbf{J}_{j}=diag[J_{j}^{\mathrm{a}},J_{j}^{\mathrm{b}}]$
are defined by Pauli matrices $\sigma_{x},\sigma_{y}$ and
$\sigma_{z}\mathbf{,}$ on-site energy differences
$\varepsilon_{j}\equiv\varepsilon_{0},$ coupling strength between
$a$-chain and $b$-chain $G_{j}\equiv G_{0}$ and hopping strength
$J_{j}^{\mathrm{a}}=J_{j}^{\mathrm{b}}\equiv J_{0}.$ Here, $N$ is
the site number of the $a$-($b$-)chain. To demonstrate the effect of
the topology configuration of such system, we will consider two
kinds of boundary conditions, which are the M\"{o}bius boundary
condition characterized as \begin{equation}
\mathbf{A}_{j+N}=\sigma_{x}\mathbf{A}_{j}\label{2-1-2}\end{equation}
 (see Fig.~\ref{fig:fig1}(a)) and the generic one characterized
as \begin{equation}
\mathbf{A}_{j+N}=\mathbf{A}_{j}\label{2-1-3}\end{equation}
 (see Fig.~\ref{fig:fig1}(b)).

The boundary condition apparently affects the system globally, and
different boundary conditions result in different symmetries. The
generic boundary condition corresponds to rotational symmetry, which
implies the generic ladder possesses $S^{1}$ topological
configuration. In contrast, the M\"{o}bius ladder is considered as a
non-orientable manifold, whose edge defines a two-point bundle over
$S^{1}$ and thus $Z^{2}$ topological configuration~\cite{DG}. This
unusual topology can induce some novel effects such as induced gauge
field and the cut-off of the electrons transmission
spectrum~\cite{Zhao}. In our paper, the $Z^{2}$ topological
configuration will contribute to the formation of the nontrivial
localized states.

\bigskip{}
%
\begin{figure}

\begin{centering}
\includegraphics[bb=22 546 569 775,clip,width=3.5in]{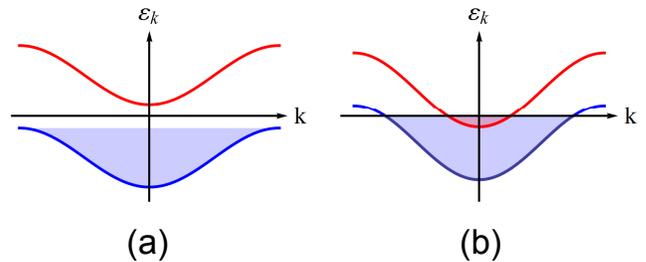}
\par\end{centering}

\caption{(Color online) Schematic illustrations of electron filling for the
monovalent ladder system before Peierls transitions, which include
(a) insulating case and (b) metallic case.}

\label{fig:fig2}
\end{figure}


By diagonalizing the above tight binding model (\ref{2-1}), two
energy bands\begin{equation}
E_{k}=\pm\sqrt{\varepsilon_{0}^{2}+G_{0}^{2}}+J_{0}\cos\left(k+\frac{\pi}{N}\right)\label{2-2-1}\end{equation}
can be obtained for the M\"{o}bius ladder, and for the generic
ladder the corresponding two energy bands are\begin{equation}
E_{k}=\pm\sqrt{\varepsilon_{0}^{2}+G_{0}^{2}}+J_{0}\cos
k.\label{2-2-2}\end{equation} The quantity of the energy shift due
to the $\pi/N$ phase shift in the energy spectrum (Eq.
(\ref{2-2-1})) of the M\"{o}bius ladder depends on the position of
the level and results from its nontrivial topology~\cite{Zhao}. If
we only consider the monovalent case that $2N$ electrons are filled
in the all negative levels for the ladder system, there are only two
kinds of energy spectra and the corresponding filling configurations
(see Fig.~\ref{fig:fig2}) before the Peierls transitions. One case
is that the valence band is entirely filled by the electrons and the
conduction band is empty when
$2J_{0}<\sqrt{\varepsilon_{0}^{2}+G_{0}^{2}},$ which corresponds to
the insulator or the semi-conductor phase (Fig.~\ref{fig:fig2}(a)).
Another case is that the electrons fill part of the conduction band
when $2J_{0}\geq\sqrt{\varepsilon_{0}^{2}+G_{0}^{2}},$ which
corresponds to the metal phase (Fig.~\ref{fig:fig2}(b)). Since the
Peierls transitions discussed below actually change the phases of
the system from conductor to insulator, only the second case is
taken into account in the following discussion.

\begin{figure}[ptb]

\begin{centering}
\includegraphics[bb=34bp 401bp 567bp 782bp,clip,width=3.5in]{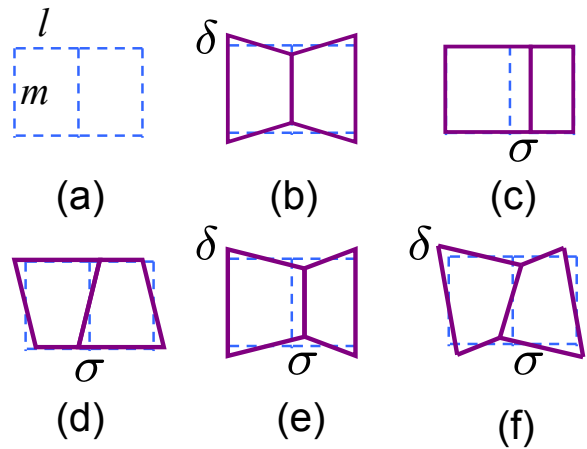}
\par\end{centering}

\caption{(Color online)Schematic illustration of the dimerization patterns
including (a) original lattice, (b) transverse, (c) columnar, (d)
staggered, (e) vertical saw-tooth, and (f) inclined saw-tooth. $m$
and $l$ are the lengths of transverse and longitudinal directions,
respectively. $\delta$ and $\sigma$ denote the static deformations
along transverse and longitudinal directions, respectively.}

\label{fig:fig3}
\end{figure}

In order to consider Peierls transition induced by electron-phonon
interaction in the ladder system, we use the Born-Oppenheimer
approximation by presuming the transverse and the longitudinal
lattice deformation (see Fig.~\ref{fig:fig3}) depicted by two
displacements $\delta$ and $\sigma,$ which are small comparing with
the lattice constant of the transverse direction $l$ and the
longitudinal direction $m.$ In Fig.~\ref{fig:fig3}, we have assumed
that the lattice is uniquely dimerized, where in general cases the
deformations depend on the locations of the sites. The above
approximation is valid because the frequency of the phonon is much
smaller than the frequency of the electrons. In the sense of the
Born-Oppenheimer approximation, we can fix the displacements of the
lattice to solve the eigenvalues of the electrons, which eventually
act as the effective potential onto the phonons.

Since the transition matrices $\mathbf{M}_{j},$ $\mathbf{J}_{j}$ of
the electrons in Eq. (\ref{2-1}) depends on the distance between the
nearest neighbour sites, after the lattice deformation the
transition matrices depend on the displacements $\delta$ and
$\sigma$ as well. Additionally, the lattice deformation is modeled
as $2N$ coupled harmonic oscillators with the Hamiltonian
\begin{eqnarray}
H_{\mathrm{p}} & = & \sum_{i=a,b}\sum_{j=0}^{N-1}\frac{K_{\mathrm{l}}}{2}\left(l_{i,j+1}-l_{i,j}\right)^{2}+\sum_{i=a,b}\sum_{j=0}^{N-1}\frac{M}{2}\left(\overset{\cdot}{l_{i,j}}\right)^{2}\notag\\
 &  & \sum_{j=0}^{N-1}\frac{K_{\mathrm{t}}}{2}(m_{a,j}-m_{b,j})^{2}+\sum_{i=a,b}\sum_{j=0}^{N-1}\frac{M}{2}\left(\overset{\cdot}{m_{i,j}}\right)^{2},\label{2-2}\end{eqnarray}
where $l_{i,j}$ and $m_{i,j}$ $(i=a,b)$ are the displacements of
the $j$-th site of $a$-chain or $b$-chain along the longitudinal
and the transverse direction, respectively. Here, $K_{\mathrm{t}}$
and $K_{\mathrm{l}}$ are spring constants of the transverse and longitudinal
directions, respectively.

The Peierls transition happens when the decrement of the total electrons
energy compensates the increment of phonon energy caused by the lattice
deformation. Here, the Fermi surface plays an important role. Actually,
after the lattice is modulated, the gaps, which are opened up at the
Fermi surface, mainly result in the decrement of the total electrons
energy.

%


All five uniform dimerization patterns ,including three simple dimerization
patterns and two hybrid ones, are presumed for manovalent case, which
is illustrated by a deformed two-square section of the ladder in Fig.~\ref{fig:fig3}.
The undimerized lattice is present in Fig.~\ref{fig:fig3}(a). The
first simple case is the rung dimerization (Fig.~\ref{fig:fig3}(b))
which possesses lattice deformation only along the transverse direction.
Along the longitudinal direction there are two different dimerization
patterns: the columnar dimerization (Fig.~\ref{fig:fig3}(c)) and
the staggered dimerization (Fig.~\ref{fig:fig3}(d)), which correspond
to same or different Peierls phases in the $a-$ and the $b-$ chain.
Last two hybrid dimerization patterns, the vertical saw-tooth (Fig.~\ref{fig:fig3}(e))
and the inclined saw-tooth (Fig.~\ref{fig:fig3}(f)), are regarded
as that the system is transversely and longitudinally dimerized simultaneously,
which possess either columnar dimerization or staggered dimerization
along longitudinal direction.

\subsection{\label{subsec:2-2}Band spectral structures of the M\"{o}bius ladder }

Since the Fourier transformation is no longer valid when the boundary
is not periodic, we introduce a new operator-value vectors $\mathbf{B}_{j}=U_{j}\mathbf{A}_{j},$
where the site-dependent unitary transformation~\cite{Zhao} is defined
as\begin{equation}
U_{j}=\frac{1}{\sqrt{2}}\left[\begin{array}{cc}
\exp(-i\theta jl) & -\exp(-i\theta jl)\\
1 & 1\end{array}\right],\label{2-3}\end{equation} where
$\theta=\pi/\left(Nl\right)$ is half of the momentum quanta. Through
this transformation, the period boundary condition
$\mathbf{B}_{j+N}=\mathbf{B}_{j}$ is retrieved for the new
operator-value vectors. In the new representation, the Hamiltonian
\begin{equation}
H_{e}=\sum_{j=0}^{N-1}\mathbf{B}_{j}^{\dag}\mathbf{M}_{j}^{\prime}\mathbf{B}_{j}-\sum_{j=0}^{N-1}\mathbf{J}_{j}^{\prime}\mathbf{B}_{j}^{\dag}\mathbf{B}_{j+1}+h.c.\label{2-4}\end{equation}
is unitarily transformed from Eq.~(\ref{2-1}), where the new
transition matrices\begin{equation}
\mathbf{M}_{j}^{\prime}=U_{j}\mathbf{M}_{j}U_{j}^{\dag}=\left[\begin{array}{cc}
G_{j} & \varepsilon_{j}\exp(-i\theta jl)\\
\varepsilon_{j}\exp(i\theta jl) & -G_{j}\end{array}\right]\label{2-5}\end{equation}
and \begin{equation}
\mathbf{J}_{j}^{\prime}=U_{j}\mathbf{J}_{j}U_{j+1}^{\dag}=\left[\begin{array}{cc}
J_{j}^{\mathrm{a}}\exp(i\theta l) & 0\\
0 & J_{j}^{\mathrm{b}}\end{array}\right]\label{2-6}\end{equation}
differ from the original ones. Such difference is considered as the
global effect induced by the M\"{o}bius boundary condition, where an
induced gauge field takes responsibility for the cutoff of the
transmission spectrum and a stark shift occurs in the energy
spectrum~\cite{Zhao}. However, this two effects actually are not
significant when we only consider electrons filling in the energy
bands for static dimerization with very large site number.

To obtain the static lattice deformations, it is necessary to minimize
the total energy $E=E_{\mathrm{e}}+E_{\mathrm{p}}$ versus the lattice
deformation. Here, based on Born-Oppenheimer approximation, the total
energy including the electron part $E_{\mathrm{e}}$ and the phonon
part $E_{\mathrm{p}}$ is obtained by diagonalizing the electron Hamiltonian
Eq. (\ref{2-4}) and the phonon Hamiltonian Eq. (\ref{2-2}), respectively.

We take the staggered dimeriation of the M\"{o}bius ladder as an
example. Let the $m$ and $l$ be the lattice constants along the
transverse and the longitudinal directions, respectively, and we can
define the static uniform deformations
$u_{j}=\left(-1\right)^{j}\delta$ and
$v_{j}=\left(-1\right)^{j}\sigma$. Because the lattice deformation
changes the coupling strength and the hopping strength from $G_{0}$
and $J_{0}$ to $G_{0}+\beta(m_{j}^{\mathrm{a}}-m_{j}^{\mathrm{b}})$
and $J_{0}+\alpha(l_{j}-l_{j+1}),$ where $\alpha$ and $\beta$ are
the rate of the changes of the longitudinal and the transverse
hopping. For the staggered dimeriation, the longitudinal lattice
deformation is $l_{j}=l+(-1)^{i+j}\sigma$ and the transverse lattice
deformation is \begin{equation}
m_{j}^{\mathrm{a}}-m_{j}^{\mathrm{b}}=\sqrt{m^{2}+(2\sigma)^{2}}-m\approx2\sigma^{2}/m.\label{2-7}\end{equation}

Therefore, with the modified coupling strength $G_{j}\equiv
G_{0}^{\prime}=G_{0}+2\beta\sigma^{2}/m$ and hopping strength
$J_{j}^{\mathrm{a}}=J_{0}+\Delta J\left(-1\right)^{j}$ for $a$-chain
and $J_{j}^{\mathrm{b}}=J_{0}+\Delta J\left(-1\right)^{j+1}$ for
$b$-chain with $\Delta J=2\alpha\sigma$, four separate energy bands
of the electrons can be obtained by diagonalizing the Hamiltonian
Eq. (\ref{2-4}) in the momentum space as \begin{subequations}
\begin{eqnarray}
\varepsilon_{j}(k) & = & \left(-1\right)^{j}\sqrt{\mu(\sigma)+(-1)^{\left\lfloor \frac{j}{2}\right\rfloor }\nu(\sigma)},\label{2-8-1}\\
\mu(\sigma) & = & G_{\mathrm{m}}^{2}+\Delta J^{2}+4J_{0}^{2}\cos^{2}\left(kl+\frac{\pi}{N}\right),\label{2-8-2}\\
\nu(\sigma) & = & 2\sqrt{G_{0}^{\prime2}\Delta J^{2}+4J_{0}^{2}G_{\mathrm{m}}^{2}\cos^{2}\left(kl+\frac{\pi}{N}\right)}\label{2-8-3}\end{eqnarray}
for $j=1,2,3,4,$ where , and $\left\lfloor \frac{j}{2}\right\rfloor $
represents the integer part of of $j/2.$

%
\begin{figure}

\begin{centering}
\includegraphics[bb=21 311 557 772,clip,width=3.5in]{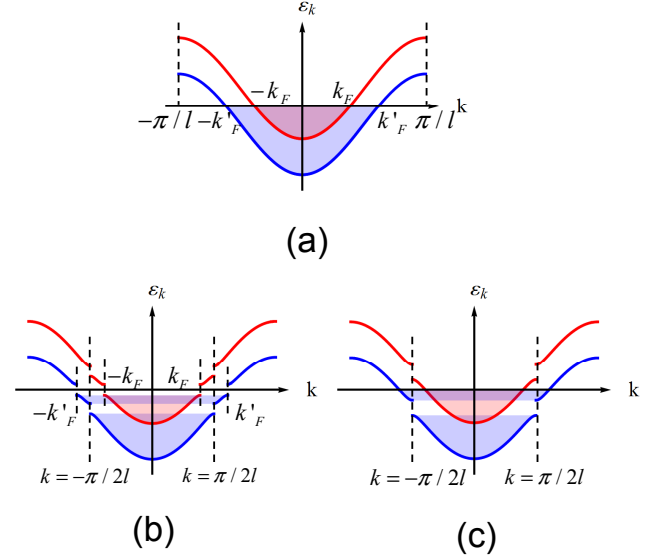}
\par\end{centering}

\caption{(Color online) Schematic spectra of ladders (a) without dimerization
(a) and (b),(c) with staggered dimerization. In (b), four gaps are
opened up at Fermi momentum $k_{f}$, $k_{f}^{\prime}$ and $k=\pm\pi/2l$,
while in (c) two gaps up at Fermi momentum disappear when the hopping
strength $J_{0}$ is sufficiently small. The shadow regions represent
the electron occupation in the energy bands.}

\label{fig:fig4}
\end{figure}


It follows the energy band diagram (Fig.~\ref{fig:fig4}(b)) that
when the hopping strength $J_{0}$ is sufficiently large, the deformation
opens four gaps in the original two overlapped bands. The two gaps
at $k=\pm\pi/2l$ are usual ones because they only arise from the
longitudinal deformation for $a$- and $b$-chain, respectively. The
other two gaps approximately locating at Fermi momentum $k_{f}=\pm\arccos(G_{\mathrm{m}}/2J_{0})/l$
in the upper band and $k_{f}^{\prime}=\pi/l-k_{f}$ in the lower band
basically arise from the coupling between the $k-$ states in $a-$chain
and the $k-\pi/l-$states in $b$-chain with strength $\alpha\sigma G_{0}^{\prime}/2J_{0}$
approximately. It is essential to indicate that when the the hopping
strength $J_{0}$ is sufficiently small, there are no gaps opened
up at Fermi surface even there exists the coupling between the $k$-states
in $a$-chain and the $k-\pi/l$-states in $b$-chain (Fig.~\ref{fig:fig4}(c)).
Thus no Peierls transitions occur.

To compare with the result of the M\"{o}bius ladder, we also
consider the dimerization in a generic ladder shown in
Fig.~\ref{fig:fig1}(b). Here, the Fourier transformation is applied
to diagonalize the electron Hamiltonian without introducing
site-dependent unitary transformation. The staggered dimerization is
still taken as an example for the generic ladder, which also
possesses four separated energy bands \end{subequations}
\begin{subequations} \begin{eqnarray}
\varepsilon_{j}^{\prime}(k) & = & \left(-1\right)^{j}\sqrt{\mu^{\prime}(\sigma)+(-1)^{\left\lfloor \frac{j}{2}\right\rfloor }\nu^{\prime}(\sigma)},\label{2-11-1}\\
\mu^{\prime}(\sigma) & = & G_{m}^{2}+\Delta J^{2}+4J_{0}^{2}\cos^{2}\left(kl\right),\label{2-11-2}\\
\nu^{\prime}(\sigma) & = & 2\sqrt{\varepsilon_{0}^{2}\Delta
J^{2}+4J_{0}^{2}G_{m}^{2}\cos^{2}\left(kl\right)},\label{2-11-3}\end{eqnarray}
which also follows the energy band diagram Fig.~\ref{fig:fig4}(b)
and Fig.~\ref{fig:fig4}(c). However, the gaps opened up at Fermi
surface are different from the ones of M\"{o}bius case, which
results in different energy of each dimerization patterns of generic
ladder from the ones of M\"{o}bius case and thus the different phase
diagrams.

\subsection{\label{subsec:2-3}Phase diagram of the M\"{o}bius ladder and the generic
ladder}

Now we focus on the case that the gaps opened up at Fermi surface
may decrease the energy of the electrons by $\Delta
E_{\mathrm{e}}=E(\delta)-E(0),$ where \end{subequations}
\begin{equation}
E(\delta)=\sum_{i=1,3}\int\varepsilon_{j}(k)dk\label{2-9}\end{equation}
and $E(0)$ is the energy without dimerization. The lattice
deformation also increases the energy of phonons by \begin{equation}
\Delta
E_{p}=4K_{l}N\sigma^{2}+2K_{t}N\frac{\sigma^{4}}{m^{2}}.\label{2-10}\end{equation}
The total energy shift $\Delta E=\Delta E_{e}+\Delta E_{p}$ versus
lattice deformation $\sigma$ is plotted in Fig.~\ref{fig:fig5}.
There is a minimum of $\Delta E$ at
$\sigma=\sigma_{s}\left(K_{l},K_{t}\right)$, which corresponds to
the stable configuration of the system. Obviously, as the order
parameter of the staggered Peierls phase transition in M\"{o}bius
ladder, $\sigma_{s}\left(K_{l},K_{t}\right)=0$ means the lattice is
not deformed corresponding to the original metal phase, while
$\sigma_{s}\left(K_{l},K_{t}\right)\neq0$ means the lattice is
spontaneously modulated to form an insulator phase. In this sense,
when all the possible $K_{t}$ and $K_{l}$ are chosen to determine
respective stable configurations, we obtain the phase diagram of the
staggered dimerization in the M\"{o}bius ladder.

The above calculation is carried out for the staggered case.
Repeating it for all deformations (Fig.~\ref{fig:fig2}) gives the
total Peierls phase diagram for the M\"{o}bius boundary condition
(Fig. ~\ref{fig:fig5}(a)). Here, the parameters are chosen as
$G_{0}=15\varepsilon_{0},J_{0}=10\varepsilon_{0},\alpha=\beta=\varepsilon_{0}/m,$
and $l=m.$

With these parameters, only three dimerization patterns survive,
which are the rung, the staggered, and the inclined saw-toothed
dimerization patterns (all are denoted by capital letters
\textquotedbl{}R\textquotedbl{}, \textquotedbl{}S\textquotedbl{},
and \textquotedbl{}I\textquotedbl{} in Fig.~\ref{fig:fig5}(a)). The
three Peierls phases occur at different regions at $K_{l}$ and
$K_{t}$. The rung dimerization occurs when $K_{l}\gg K_{t},$ the
staggered dimerization occurs when $K_{l}\ll K_{t},$ and the
inclined saw-toothed dimerization occurs when $K_{l}\approx K_{t}$
is sufficiently small. When $K_{l}\approx K_{t}$ is sufficiently
large, there is no dimerization emerging in the M\"{o}bius ladder.

%
\begin{figure}[ptb]

\begin{centering}
\includegraphics[bb=14bp 515bp 557bp 793bp,clip,width=3.5in]{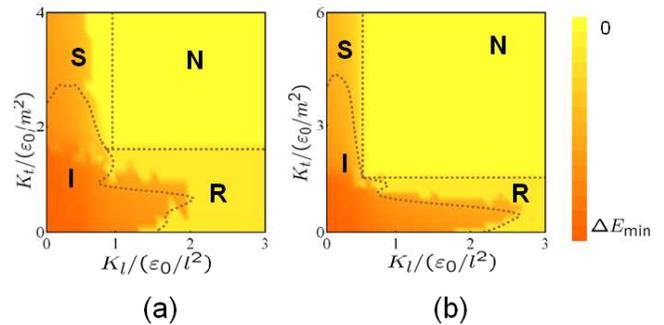}
\par\end{centering}

\caption{(Color online) The phase diagrams of the ladder system with
(a) the generic and (b) the M\"{o}bius boundary conditions, which
are plotted versus $\left(K_{l},K_{t}\right)$ in (a) and (b),
respectively. The parameters are chosen as
$G_{0}=15\varepsilon_{0}$, $J_{0}=10\varepsilon_{0}$,
$\alpha=\beta=\varepsilon_{0}/m$, and $l=m$$.$The distribution of
the total energy $\Delta E$ versus $\left(K_{l},K_{t}\right)$
determines the boundaries of the phases, which are plotted as dashed
lines. Here S, I, R, and N represent the staggered, the inclined
saw-tooth, rung and no dimerization, respectively.}

\label{fig:fig5}
\end{figure}


With the same procedure, by minimizing $\Delta E$ of the generic
ladder, we obtain the total Peierls phase diagram for the generic
boundary condition (Fig.~\ref{fig:fig5}(b)). Notice here that the
energy bands completely filled with electrons become $\varepsilon_{j}^{\prime}(k)$
$(j=1,3)$ in Eq. (\ref{2-11-1}).

With the same parameters of the M\"{o}bius case, the basic
properties of the Peierls phases of generic ladder is similar to the
one of M\"{o}bius case. However, the region of the staggered
dimerization pattern under the M\"{o}bius boundary condition shrinks
comparing with the generic one. This fact means that the metal phase
is preferable for a M\"{o}bius ladder system. Therefore, the above
phase diagrams show that the conducting properties can be
dramatically changed in when the topology of the ladder is switched.
Because the inclined saw-toothed phase contains the staggered
dimerization along longitudinal direction, it is changed the same
way as the staggered one. The Peierls phase for rung dimerization is
exactly the same whatever the boundary condition is. Although the
columnar and the vertical saw-toothed dimerization do not appear in
the current phase diagrams in Fig.~\ref{fig:fig5}, their
corresponding Peierls phases are the same for different boundary
conditions of the ladder system when $G_{0}\ll\varepsilon_{0}$.
Therefore, to further consider the topological effect on conducting
properties, only the staggered dimerization pattern is taken into
account and the existence of the localized states will be revisited
for our Q1D system.

\section{\label{sec:three}Continuum Model and Solitonary Solutions}

\subsection{localized states in the M\"{o}bius ladder}

We adapt the continuous field approach by regarding the M\"{o}bius
ladder as a one-dimensional system with long range hopping
(Fig.~\ref{fig:fig1}(c)). The detailed derivation of the continuum
model from the lattice Hamiltonian Eq.~(\ref{2-1}) and (\ref{2-2})
is presented in App.~\ref{app:one}. Without loss of the generality,
we focus on the special case with $\varepsilon_{0}=0.$ For a
continuous field approach, it is crucial to introduce an order
parameter \begin{equation}
\Delta(x)=-4\alpha\phi(x),\label{3=0}\end{equation} where $\phi(x)$
is the continuous limit of $\phi_{j}=\left(-1\right)^{j}u_{j}.$

The Hamiltonian of the continuum model $H=H_{\mathrm{e}}+H_{\mathrm{p}}$
contains the the phonon part \begin{eqnarray}
H_{\mathrm{p}} & = & \int_{-L}^{L}dx\left\{ \frac{K_{\mathrm{l}}}{8\alpha^{2}l}\Delta^{2}(x)+\frac{M}{32\alpha^{2}l}\overset{\cdot}{\Delta^{2}}(x)\right.+\notag\\
 &  & \left.\frac{K_{\mathrm{t}}}{4^{6}m^{2}\alpha^{4}l}\left[\Delta(x+L)-\Delta(x)\right]^{4}\right\} \label{3-1}\end{eqnarray}
and the electron part \begin{equation}
H_{\mathrm{e}}=\int_{-L}^{0}dx\Phi(x)\mathscr{H}_{\mathrm{m}}\Phi(x),\label{3-2}\end{equation}
where $M$ is the mass of the particle and $L=Nl$ is the length of
the $a$-chain ($b$-chain), which approaches infinity at the end of
the calculation. Here, the subindex $\mathrm{m}$ stands for the
M\"{o}bius ladder. In the electron part, the hopping electron could
be described with a 4-component spinor
$\Phi(x)=\left[\begin{array}{cccc} \phi_{1}(x) & \phi_{2}(x) &
\phi_{3}(x) & \phi_{4}(x)\end{array}\right]^{\mathrm{T}}.$
Physically, $\phi_{1}(x)$($\phi_{3}(x)$) and
$\phi_{2}(x)$($\phi_{4}(x)$) respectively represent the
left-traveling wave and right-traveling wave in $a$-chain
($b$-chain). In this spinor representation, the Hamiltonian density
is expressed by Pauli matrices $\sigma_{x},$ $\sigma_{y},$
$\sigma_{z}$ as\begin{equation}
\mathscr{H}_{\mathrm{m}}=\left[\begin{array}{cc}
iv_{\mathrm{f}}\sigma_{z}\partial_{x}+\Delta(x)\sigma_{x} & G\left(x\right)\\
G\left(x\right) & iv_{\mathrm{f}}\sigma_{z}\partial_{x}+\Delta(x+L)\sigma_{x}\end{array}\right]\label{3-3}\end{equation}
where $v_{\mathrm{f}}=2lJ_{0},$ $\sigma_{z},\sigma_{x}$ are Pauli
matrices, and \begin{equation}
G\left(x\right)=G_{0}+\frac{\beta}{32m\alpha^{2}}\left[\Delta(x)-\Delta(x+L)\right]^{2}\label{3-4}\end{equation}
is effective coupling between the $a$-chain and $b$-chain.

To reflect the M\"{o}bius boundary condition in our Q1D model
(Fig.~\ref{fig:fig1}(c)), we take the period $2L$ for boundary
conditions \begin{equation}
\Phi\left(x+2L\right)=\Phi\left(x\right)\label{3-4-1}\end{equation}
 rather than $L$ for the generic case. With this boundary condition,
we solve the Bogoliubov-de Gennes (BdG) equation \begin{equation}
\mathscr{H}_{\mathrm{m}}\Phi_{i}\left(x\right)=\varepsilon_{i}\Phi_{i}\left(x\right),\label{3-5}\end{equation}
where $i$ represents the $i$-th energy band of the spectrum and
$\Phi_{i}\left(x\right)=[\phi_{1}^{i}\left(x\right),\phi_{2}^{i}\left(x\right),\phi_{3}^{i}\left(x\right),\phi_{4}^{i}\left(x\right)]^{\mathrm{T}}$.
At zero temperature, the order parameter $\Delta\left(x\right)$ satisfies
the self-consistent equations \begin{eqnarray}
 &  & \frac{K_{\mathrm{l}}}{4\alpha^{2}l}\Delta(x)-\frac{K_{\mathrm{t}}}{4^{5}m^{2}\alpha^{4}l}(\Delta(x+L)-\Delta(x))^{3}\notag\\
 & = & \left\{ \begin{array}{c}
\sum_{i}2\mathbf{Re}\left[\phi_{1}^{i,\ast}\left(x\right)\phi_{2}^{i}\left(x\right)\right],\text{for }x\leq0,\\
\sum_{i}2\mathbf{Re}\left[\phi_{3}^{i,\ast}\left(x\right)\phi_{4}^{i}\left(x\right)\right],\text{for }x>0,\end{array}\right.\label{3-6}\end{eqnarray}
which is obtained by the functional variation of \begin{equation}
E\left(\Delta\left(x\right)\right)=\sum_{i}\varepsilon_{i}+H_{\mathrm{p}}\label{3-6-1}\end{equation}
 with respect to $\delta\Delta\left(x\right)$ and $\delta\Delta\left(x+L\right).$
The sum is over the energy levels below the Fermi surface. In principle,
the eigenenergies $\varepsilon_{i},$ the eigenfunctions $\Phi_{i}\left(x\right)$
and the order parameters $\Delta(x)$ and $\Delta\left(x+L\right)$
can be completely determined by the BdG equation in Eq. (\ref{3-5})
and the self-consistent equation in in Eq. (\ref{3-6}).

After introducing new 4-component spinor $\Psi_{i}(x)=\left[\begin{array}{cccc}
\varphi_{1}^{i}(x) & \varphi_{2}^{i}(x) & \varphi_{3}^{i}(x) & \varphi_{4}^{i}(x)\end{array}\right]^{\mathrm{T}}$ by $\Psi_{i}(x)=U\Phi_{i}\left(x\right)$, where the unitary matrix
is defined as \begin{equation}
U=\frac{1}{\sqrt{2}}\left[\begin{array}{cccc}
1 & i & 0 & 0\\
1 & -i & 0 & 0\\
0 & 0 & 1 & i\\
0 & 0 & 1 & -i\end{array}\right],\label{3-7}\end{equation}
the BdG equation in Eq. (\ref{3-5}) can be simplified as \begin{equation}
\mathscr{H}_{\mathrm{m}}^{\prime}\Psi_{i}(x)=\varepsilon_{i}\Phi_{i}\left(x\right)\label{3-7-1}\end{equation}
 with new Hamiltonian density \begin{equation}
\mathscr{H}_{\mathrm{m}}^{\prime}=\left[\begin{array}{cc}
iv_{\mathrm{f}}\sigma_{x}\partial_{x}-\Delta(x)\sigma_{y} & G_{0}\\
G_{0} & iv_{\mathrm{f}}\sigma_{x}\partial_{x}-\Delta(x+L)\sigma_{y}\end{array}\right].\label{3-8}\end{equation}

As we shown as follows, some solutions of the above BdG equation can
exist as localized states. In the following, we only consider the
case $K_{l}\ll K_{t}$. In this case, three dimerization patterns of
rung (Fig.~\ref{fig:fig3}(b)), vertical-saw tooth
(Fig.~\ref{fig:fig3}(e)) and inclined saw-tooth
(Fig.~\ref{fig:fig3}(f)) occur rarely. Thus we only need to compare
the energy of staggered dimerization with columnar one. Here, we
revisit M\"{o}bius ladder system with the staggered dimerization
characterized by \begin{equation}
\Delta\left(x\right)=-\Delta\left(x+L\right).\label{3-8-1}\end{equation}
 In this phase, the order parameters in $a$-chain and $b$-chains
are opposite and display a Peierls phases domain wall when the site
number $N$ is even. We also notice that the columnar dimerization
to be compared is characterized by \begin{equation}
\Delta\left(x\right)=\Delta\left(x+L\right).\label{3-8-2}\end{equation}
 For the staggered case, we assume a kink deformation as the form\begin{equation}
\Delta\left(x\right)=\Delta\tanh(x/\xi)\label{3-8-3}\end{equation}
 with $\xi=v_{\mathrm{f}}/\Delta$, which is so small that the effective
coupling between the $a$-chain and $b$-chain $G\left(x\right)\approx G_{0}.$

To solve the new BdG equation, some symmetries of the Hamiltonian
density can be used to simplify the calculation. The Hamiltonian density
actually possesses the discrete symmetry \begin{equation}
W^{\dag}\mathscr{H}_{\mathrm{m}}^{\prime}W=\mathscr{H}_{\mathrm{m}}^{\prime}\label{3-9}\end{equation}
with an anti-diagonal matrix \begin{equation}
W=W^{\dag}=\left[\begin{array}{cccc}
0 & 0 & 0 & 1\\
0 & 0 & 1 & 0\\
0 & 1 & 0 & 0\\
1 & 0 & 0 & 0\end{array}\right]\label{3-10}\end{equation}
denoting the mirror reflection transformation. This symmetry guarantees
that if $\Psi_{i}(x)$ is an eigen function of $\mathscr{H}_{\mathrm{m}}^{\prime}$
with eigenenergy $\varepsilon_{i}$, the $W^{\dag}\Psi_{i}(x)$ is
also the eigen function of $\mathscr{H}_{\mathrm{m}}^{\prime}$ with
the same eigenenergy $\varepsilon_{i}.$ Obviously, the eigenvalues
of matrix W are $\pm1$. Together with the translational symmetry
characterized by momentum quantum numbers, the total Hilbert space
can be spanned by these bases. Therefore, the eigen function $\Psi_{i}(x)$
either has the form \begin{equation}
\Psi_{i}(x)=W\Psi_{i}(x)\label{3-11-1}\end{equation}
 or \begin{equation}
\Psi_{i}(x)=-W\Psi_{i}(x).\label{3-11-2}\end{equation}

In this sense, as the solutions of the BdG equation with energy $\varepsilon_{\mathrm{s}}=0$,
two degenerate solitonary states can be found as one with non-vanishing
components \begin{equation}
\varphi_{2}^{\mathrm{s}}\left(x\right)=\varphi_{3}^{\mathrm{s}}\left(x\right)=F_{+}^{\mathrm{s}}(x),\label{3-11-3}\end{equation}
and another with non-vanishing components \begin{equation}
\varphi_{2}^{\mathrm{s}}\left(x\right)=-\varphi_{3}^{\mathrm{s}}\left(x\right)=F_{-}^{\mathrm{s}}(x)\label{3-11-4}\end{equation}
 for \begin{equation}
F_{\pm}^{\mathrm{s}}\left(x\right)=\sqrt{\frac{1}{2\xi}}\exp\left(\pm i\frac{G_{0}}{v_{\mathrm{f}}}x\right)\text{sech}\left(\frac{x}{\xi}\right),\label{3-12}\end{equation}
where the subindex $\mathrm{s}$ denotes solitonary solutions. These
solitonary states are the localized states located at the midgap.
Since there is no such solitonary state in the generic ladder, the
existence of the solitons is absolutely topological effect.

\begin{figure}

\begin{centering}
\includegraphics[bb=28bp 555bp 570bp 779bp,clip,width=3.5in]{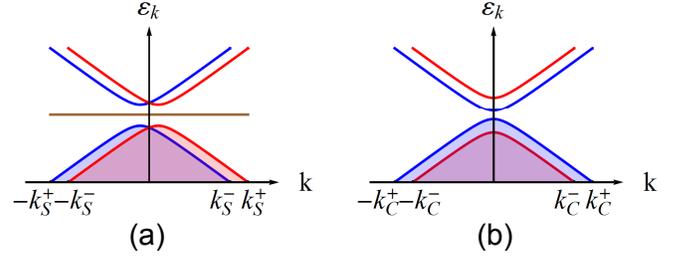}
\par\end{centering}

\caption{(color online) Schematics of the energy spectra of the valence bands
under (a) staggered and (b) columnar dimerizations, where $k_{\mathrm{S}}^{\pm}=k_{\mathrm{f}}\mp G_{0}/v_{\mathrm{f}}$
and $k_{\mathrm{C}}^{\pm}=\sqrt{G_{0}^{2}+v_{\mathrm{f}}^{2}k_{\mathrm{f}}^{2}\pm2G_{0}\sqrt{v_{\mathrm{f}}^{2}k_{\mathrm{f}}^{2}+\Delta^{2}}}/v_{\mathrm{f}}.$
The shadow regions represent the electron occupation in the energy
bands, and the brown straight line represents the solitonary states.}

\label{fig:fig6}
\end{figure}

We note that the another two bands \begin{equation}
\varepsilon_{\mathrm{v}}^{\pm}=-\sqrt{\left(v_{\mathrm{f}}k\pm G_{0}\right)^{2}+\Delta}\label{3-13}\end{equation}
(illustrated in Fig.~\ref{fig:fig6}(a) as two overlapped shadowed
domains) fully occupied by the electrons correspond to eigen functions
\begin{subequations} \begin{eqnarray}
\varphi_{1}^{\mathrm{v},\pm}\left(x\right) & = & \pm\varphi_{4}^{\mathrm{v},\pm}\left(x\right)=\frac{i}{2\sqrt{L}}e^{-ikx},\label{3-14-1}\\
\varphi_{2}^{\mathrm{v},\pm}\left(x\right) & = & \pm\varphi_{3}^{\mathrm{v},\pm}\left(x\right)=\frac{1}{2\sqrt{L}}F_{\pm}^{\mathrm{v}}(x)e^{-ikx},\label{3-14-2}\end{eqnarray}
where \end{subequations} \begin{equation}
F_{\pm}^{\mathrm{v}}(x)=\frac{\Delta}{\varepsilon_{\mathrm{v}}^{\pm}}\left[\tanh\frac{\Delta x}{v_{\mathrm{f}}}+i\frac{\left(v_{\mathrm{f}}k\pm G_{0}\right)}{\Delta}\right]\label{3-15}\end{equation}
represents a deviation from a plane wave in the kink order. Here,
the subindex $\mathrm{v}$ stands for the valence bands below the
Fermi surface.

Then, it follows from the self-consistent equations Eq. (\ref{3-6})
that

\begin{eqnarray}
 &  & \frac{K_{l}}{4\alpha^{2}l}+\frac{2K_{t}}{4^{4}m^{2}\alpha^{4}l}\Delta^{2}\tanh^{2}(x/\xi)\notag\\
 & = & \frac{1}{2\pi}\int_{-k_{\mathrm{f}}}^{k_{\mathrm{f}}}\frac{dk}{\sqrt{v_{\mathrm{f}}^{2}k_{\mathrm{f}}^{2}+\Delta^{2}}}.\label{3-16}\end{eqnarray}
Usually, the solitons are localized around the original point with
the width $\xi$ of several lattice constants, which is much smaller
than the total length of the ladder system. For most sites far away
from the original point as $|x|\gg |\xi|$, we assume
$\tanh^{2}(x/\xi)\approx1$. Since the second term is small comparing
with the first term at the left side of the above equation, we
obtain approximate order parameter

\begin{equation}
\Delta\approx\Delta_{0}\exp(-B\Delta_{0}^{2}),\label{3-17}\end{equation}
where $\Delta_{0}=W\exp(-A)$ is the order parameter for one dimensional
uniformly dimerized system, $W=2v_{\mathrm{f}}k_{\mathrm{f}}$, $A=v_{\mathrm{f}}\pi K_{l}/8\alpha^{2}l$,
and $B=v_{\mathrm{f}}\pi K_{t}/2^{7}m^{2}\alpha^{4}l$ with $k_{\mathrm{f}}$
is the Fermi momentum.

Moreover, we can further prove that the above solitonary states are
the ground states. To this end we calculate the total energy of the
electron-phonon system $E_{\mathrm{T}}^{\mathrm{S}}$ according to
the phase shift~\cite{PT2} of the eigenstates. The phase shifts
are determined by the eigenstates of the band electrons$\,$\ when
$x$ tends to $\pm\infty$ as\begin{eqnarray}
\lim_{x\rightarrow\pm\infty}F_{\pm}^{\mathrm{v}}(x) & = & \frac{\Delta}{\varepsilon_{\mathrm{v}}^{\pm}}\left[\pm1+i\frac{\left(v_{\mathrm{f}}k\pm G_{0}\right)}{\Delta}\right]\notag\\
 & \propto & \exp(i\theta_{\pm\infty}^{\pm}\left(k\right)),\label{3-18}\end{eqnarray}
which reads \begin{subequations} \begin{eqnarray}
\theta_{+\infty}^{\pm}\left(k\right) & = & \arctan\left(\frac{v_{\mathrm{f}}k\pm G_{0}}{\Delta}\right),\label{3-19-1}\\
\theta_{-\infty}^{\pm}\left(k\right) & = & -\arctan\left(\frac{v_{\mathrm{f}}k\pm G_{0}}{\Delta}\right)-\pi.\label{3-19-2}\end{eqnarray}
Therefore, the total phase shift of the eigenstates is defined by
their difference as \end{subequations} \begin{equation}
\theta^{\pm}\left(k\right)=\theta_{+\infty}^{\pm}\left(k\right)-\theta_{-\infty}^{\pm}\left(k\right)=\pi+2\arctan\left(\frac{v_{\mathrm{f}}k\pm G_{0}}{\Delta}\right).\label{3-20}\end{equation}

%


A straightforward algebra explicitly gives \cite{PT2}
\begin{equation}
E_{\mathrm{T}}^{\mathrm{S}}=E_{\mathrm{T}}^{\mathrm{C}}+\frac{4\Delta}{\pi}-\frac{G_{0}^{2}}{v_{\mathrm{f}}k_{\mathrm{f}}}+\delta(K_{\mathrm{t}}),\label{3-21}\end{equation}
where $E_{\mathrm{T}}^{\mathrm{C}}$ is the total energy for the
columnar dimerization (Fig.~\ref{fig:fig3}(c)), and
$\delta(K_{\mathrm{t}})=13K_{\mathrm{t}}\Delta^{3}v_{\mathrm{f}}/(3\times4^{4}m^{2}\alpha^{4}l)$
results from the coupling between the $a$- and $b$-chain. The second
term in $E_{\mathrm{T}}^{\mathrm{S}}$ is usual energy increment due
to the existence of solitonary states. The third term in
$E_{\mathrm{T}}^{\mathrm{S}}$ results from the difference in the
total energies in two filling ways. One corresponds to the staggered
dimerization (Fig.~\ref{fig:fig6}(a)) with two lower bands
$-\sqrt{\left(v_{\mathrm{f}}k\pm G_{0}\right)^{2}+\Delta^{2}}$
occupied by electrons, while the other corresponds to columnar one
(Fig.~\ref{fig:fig6}(b)) with two lower bands $\pm
G_{0}-\sqrt{\left(v_{\mathrm{f}}k\right)^{2}+\Delta^{2}}$ occupied.
For the latter the energy of electrons increases because a part of
electrons are forced to occupy higher energy levels. If $G_{0}$ is
so large that $\delta
E=E_{\mathrm{T}}^{\mathrm{S}}-E_{\mathrm{T}}^{\mathrm{C}}$ is
negative, the ground state of the M\"{o}bius ladder system
corresponds to the staggered dimerization rather than the columnar
one. In this case the solitonary states are localized states as the
ground state.

\subsection{Comparison with the generic ladder}

It is a complete topological effect that the ground state is
localized. To demonstrate this, the continuum model for the generic
ladder $H^{\prime}=H_{\mathrm{e}}^{\prime}+H_{\mathrm{p}}^{\prime}$
is presented to compare with the M\"{o}bius case. Here, the phonon
part is \begin{eqnarray}
H_{\mathrm{p}}^{\prime} & = & \sum_{c=a,b}\int_{-\infty}^{\infty}dx\left\{ \frac{K_{\mathrm{l}}}{8\alpha^{2}l}\Delta_{\mathrm{c}}^{2}(x)+\frac{M}{32\alpha^{2}l}\overset{\cdot}{\Delta_{\mathrm{c}}^{2}}(x)\right\} +\notag\\
 &  & \int_{-\infty}^{\infty}dx\frac{K_{\mathrm{t}}}{4^{6}m^{2}\alpha^{4}l}\left[\Delta_{\mathrm{a}}(x)-\Delta_{\mathrm{b}}(x)\right]^{4}\label{3-22}\end{eqnarray}
and the electron part is \begin{equation}
H_{\mathrm{e}}^{\prime}=\int_{-\infty}^{\infty}dx\Phi^{\prime}(x)\mathscr{H}_{\mathrm{g}}\Phi^{\prime}(x),\label{3-23}\end{equation}
where the 4-component spinor $\Phi_{i}^{g}\left(x\right)$ has the
same physical meaning as the one in the M\"{o}bius case, and the
Hamiltonian density reads\begin{equation}
\mathscr{H}_{\mathrm{g}}=\left[\begin{array}{cc}
iv_{\mathrm{f}}\sigma_{z}\partial_{x}+\Delta_{\mathrm{a}}(x)\sigma_{x} & G_{0}^{\prime}\\
G_{0}^{\prime} & iv_{\mathrm{f}}\sigma_{z}\partial_{x}+\Delta_{\mathrm{b}}(x)\sigma_{x}\end{array}\right]\label{3-24}\end{equation}
where \begin{equation}
G\left(x\right)=G_{0}+\frac{\beta}{32m\alpha^{2}}\left[\Delta_{\mathrm{a}}(x)-\Delta_{\mathrm{b}}(x)\right]^{2}\label{3-25}\end{equation}
is effective coupling between the $a$-chain and $b$-chain. It is
noticed that the order parameters here are no longer unified in one
chain, and are defined as $\Delta_{\mathrm{a}}(x)$ and $\Delta_{\mathrm{b}}\left(x\right)$
for $a$-chain and $b$-chain, respectively. Additionally, the boundary
condition for the generic ladder is $\Phi^{\prime}\left(x+L\right)=\Phi^{\prime}\left(x\right)$
with period $L.$ We will show that because of the trivial topology
of the generic ladder, there is no solitonary solution for the dimerization
of the generic ladder.

At zero temperature, the order parameter $\Delta\left(x\right)$ satisfies
the self-consistent equations \begin{subequations} \begin{eqnarray}
 &  & \frac{K_{\mathrm{l}}}{4\alpha^{2}l}\Delta_{a}(x)+\frac{K_{\mathrm{t}}}{4^{5}m^{2}\alpha^{4}l}(\Delta_{\mathrm{a}}(x)-\Delta_{\mathrm{b}}(x))^{3}\notag\\
 & = & \sum_{i}2\mathbf{Re}\left[\left(\phi_{1}^{i,\prime}\left(x\right)\right)^{\ast}\phi_{2}^{i,\prime}\left(x\right)\right],\label{3-26-1}\\
 &  & \frac{K_{\mathrm{l}}}{4\alpha^{2}l}\Delta_{b}(x)-\frac{K_{\mathrm{t}}}{4^{5}m^{2}\alpha^{4}l}(\Delta_{\mathrm{a}}(x)-\Delta_{\mathrm{b}}(x))^{3}\notag\\
 & = & \sum_{i}2\mathbf{Re}\left[\left(\phi_{3}^{i,\prime}\left(x\right)\right)^{\ast}\phi_{4}^{i,\prime}\left(x\right)\right],\label{3-26-2}\end{eqnarray}
respectively, which are obtained by the functional variation of \end{subequations}
\begin{equation}
E\left(\Delta\left(x\right)\right)=\sum_{i}\varepsilon_{i}+H_{\mathrm{p}}\label{3-26-3}\end{equation}
 with respect to $\delta\Delta_{\mathrm{a}}\left(x\right)$ and $\delta\Delta_{\mathrm{b}}\left(x\right).$
The sum is over the energy levels below the Fermi surface. For the
staggered case, we assume a kink deformation $\Delta_{\mathrm{a}}\left(x\right)=-\Delta_{\mathrm{b}}\left(x\right)=\Delta.$

After applying the same transformation, the BdG equation of the staggered
dimerized generic ladder is \begin{equation}
\mathscr{H}_{\mathrm{g}}^{\prime}\Psi_{i}^{\prime}(x)=\varepsilon_{i}^{\prime}\Psi_{i}^{\prime}(x)\label{3-26-4}\end{equation}
 with new Hamiltonian density \begin{equation}
\mathscr{H}_{\mathrm{g}}^{\prime}=\left[\begin{array}{cc}
iv_{\mathrm{f}}\sigma_{x}\partial_{x}-\Delta\sigma_{y} & G_{0}^{\prime}\\
G_{0}^{\prime} & iv_{\mathrm{f}}\sigma_{x}\partial_{x}+\Delta\sigma_{y}\end{array}\right],\label{3-27}\end{equation}
where $G_{0}^{\prime}=G_{0}+\beta\Delta^{2}/8m\alpha^{2}.$

The exact spectrum solved from the BdG equations contains four energy
bands including the lower two bands $\varepsilon_{\mathrm{v}}^{\prime,\pm}=-\sqrt{\left(v_{\mathrm{f}}k\pm G_{0}\right)^{2}+\Delta}$
occupied by the electrons and the higher two bands $\varepsilon_{\mathrm{v}}^{\prime,\pm}=\sqrt{\left(v_{\mathrm{f}}k\pm G_{0}\right)^{2}+\Delta}$
without being occupied. The corresponding eigen function are \begin{subequations}
\begin{eqnarray}
\varphi_{1,\mathrm{g}}^{\mathrm{v},\pm}\left(x\right) & = & \pm\varphi_{4,\mathrm{g}}^{\mathrm{v},\pm}\left(x\right)=\mp\frac{1}{2\sqrt{L}}e^{i\left(\frac{G_{0}^{\prime}}{v_{\mathrm{f}}}-k\pm\theta_{k}^{\pm}\right)x},\label{3-28-1}\\
\varphi_{2,\mathrm{g}}^{\mathrm{v},\pm}\left(x\right) & = & \pm\varphi_{3,\mathrm{g}}^{\mathrm{v},\pm}\left(x\right)=\frac{1}{2\sqrt{L}}e^{i\left(\frac{G_{0}^{\prime}}{v_{\mathrm{f}}}-k\right)x},\label{3-28-2}\end{eqnarray}
for the two lower bands and \end{subequations} \begin{subequations}
\begin{eqnarray}
\varphi_{1,\mathrm{g}}^{\mathrm{v},\pm}\left(x\right) & = & \pm\varphi_{4,\mathrm{g}}^{\mathrm{v},\pm}\left(x\right)=\pm\frac{1}{2\sqrt{L}}e^{i\left(\frac{G_{0}^{\prime}}{v_{\mathrm{f}}}-k\pm\theta_{k}^{\pm}\right)x},\label{3-29-1}\\
\varphi_{2,\mathrm{g}}^{\mathrm{v},\pm}\left(x\right) & = &
\pm\varphi_{3,\mathrm{g}}^{\mathrm{v},\pm}\left(x\right)=\frac{1}{2\sqrt{L}}e^{i\left(\frac{G_{0}^{\prime}}{v_{\mathrm{f}}}-k\right)x},\label{3-29-2}\end{eqnarray}
\end{subequations} for the two upper bands, respectively. Here, the
subindex $\mathrm{g}$ stands for the generic ladder.

There is no solitonary solution existing for the staggered dimerization
pattern of the generic ladder. Therefore, the localized states are
the complete topological effect.

The M\"{o}bius ladder with staggered dimerization actually behaves
like a topological insulator. Naturally, the M\"{o}bius
configuration is $\mathbb{Z}^{2}$ topologically invariant and
gapless localized states exist in the gap. Aditionally, the topology
of the system can protect the solitonary states from external
perturbations. For example, when the soliton propagates along the
longitudinal directions without spreading, the energy increment
caused by moving soliton with velocity $v_{\mathrm{s}}$ from the
time evolution of order parameter
$\Delta\left(x,t\right)\equiv\Delta\tanh(x-v_{\mathrm{s}}t)/\xi$ is
$\Delta
E_{\mathrm{s}}=Mv_{\mathrm{s}}^{2}\Delta^{3}/(24v_{\mathrm{f}}\alpha^{2}l)$,
which could be much smaller than the exciting energy $\delta E$. It
indicates that the moving solitons can propagate in the M\"{o}bius
ladder without dispersion and thus is robust to external
perturbations.

\section{\label{sec:four}CONCLUSION}

We study the topological properties of Peierls transitions in a
monovalent M\"{o}bius ladder in contrast to the Peierls transitions
in a generic ladder. According to lattice deformation along the
transverse and longitudinal directions of the ladder configuration,
there exist plenty Peierls phases corresponding to various
dimerization patterns. The insulator phase resulted from staggered
modulation along longitudinal direction behaves as a topological
insulator, which is different from mundane band insulator. Actually,
this non-trivial insulator originates from the Peierls phases
boundary induced by the non-trivial $\mathbb{Z}^{2}$ topological
configuration.

\acknowledgments The authors thanks Nan Zhao for helpful discussion.
This work is supported by NSFC No.10474104, No.60433050, and
No.10704023, NFRPC No.2006CB921205 and 2005CB724508.

\appendix*

\section{\label{app:one}The Derivation of the Continuum Model}

When the site number $N$ is so large that the characteristic wave
length of the eigen function is greater than the lattice constant
$l$, the continuum field approach is appropriately adapted by
regarding the M\"{o}bius ladder as a one-dimensional system with
long range hopping (Fig.\ref{fig:fig1}(c)). We consider the upper
chain ($a$-chain) and lower chain ($b$-chain) as the first half and
second half of a whole chain with $2N$ sites, which corresponds to
the mapping

\begin{equation}
a_{j}\rightarrow A_{j},b_{j}\rightarrow A_{j+N}\label{a-1}\end{equation}
with fermionic operators $A_{j}.$ After the mapping, the electron
part of the lattice Hamiltonian\ in Eq. (\ref{2-1}) with a fixed
deformation configuration $\{u_{j}\}=\{u_{j}^{a},u_{j}^{b}\}$ can
be rewritten as\begin{eqnarray}
H_{\mathrm{e}} & = & \sum_{j=0}^{N-1}\varepsilon_{0}A_{j}^{\dagger}A_{j}-\sum_{j=0}^{N-1}J_{j}^{\mathrm{a}}\left(A_{j}^{\dagger}A_{j+1}+h.c.\right)\notag\\
 &  & -\sum_{j=N}^{2N-1}\varepsilon_{0}A_{j}^{\dagger}A_{j}-\sum_{j=N}^{2N-1}J_{j}^{\mathrm{b}}\left(A_{j}^{\dagger}A_{j+1}+h.c.\right)\notag\\
 &  & -\sum_{j=0}^{N-1}G_{j}\left(A_{j}^{\dagger}A_{j+N}+A_{j+N}^{\dagger}A_{j}\right),\label{a-2}\end{eqnarray}
where the modified coupling constants are \begin{subequations} \begin{eqnarray}
G_{j} & = & G_{0}+\beta\frac{\left(u_{j}^{\mathrm{b}}-u_{j}^{\mathrm{a}}\right)^{2}}{2m},\label{a-3-1}\\
J_{j}^{\mathrm{c}} & = & J_{0}+\alpha\left(u_{j+1}^{\mathrm{c}}-u_{j}^{\mathrm{c}}\right),\left(\mathrm{c=a,b}\right),\label{a-3-2}\end{eqnarray}
and indecies $\mathrm{a,b}$ stand for the original $a$-chain and
$b$-chain. The energy of the phonon in Eq. (\ref{2-2}) can also
be obtained as \end{subequations} \begin{eqnarray}
H_{\mathrm{p}} & = & \sum_{j=0}^{N-1}\frac{K_{\mathrm{l}}}{2}\left(u_{j+1}^{\mathrm{a}}-u_{j}^{\mathrm{a}}\right)^{2}+\sum_{j=0}^{N-1}\frac{M}{2}\left(\dot{u_{j}^{\mathrm{a}}}\right)^{2}\notag\\
 &  & +\sum_{j=N}^{2N-1}\frac{K_{\mathrm{l}}}{2}\left(u_{j+1}^{\mathrm{b}}-u_{j}^{\mathrm{b}}\right)^{2}+\sum_{j=N}^{2N-1}\frac{M}{2}\left(\dot{u_{j}^{\mathrm{b}}}\right)^{2}\notag\\
 &  & +\sum_{j=0}^{N-1}\frac{K_{\mathrm{t}}}{2}\frac{\left(u_{j}^{\mathrm{b}}-u_{j}^{\mathrm{a}}\right)^{4}}{4m^{2}}.\label{a-3}\end{eqnarray}

Usually the wavefunction varies greatly from site to site under dimerization,
which means the coordinate $j$ is not suitable for the continuous
field approach. However, if we introduce the new coordinate \begin{equation}
x_{j}\rightarrow(2j+\frac{1}{2})l\label{a-4}\end{equation}
as the center of the $2j$-th and $\left(2j+1\right)$-th sites, the
wavefunction varies slowly and the continuous field approach is valid.

In this sense, the new fermionic field operators\begin{equation}
\begin{cases}
\varphi_{1}\left(x_{j}\right)=\frac{1}{\sqrt{2}}\left(-1\right)^{j}\left(iA_{2j}+A_{2j+1}\right),\\
\varphi_{2}\left(x_{j}\right)=\frac{1}{\sqrt{2}}\left(-1\right)^{j}\left(A_{2j}+iA_{2j+1}\right),\end{cases}\label{a-5}\end{equation}
which satisfy the anti-commutate relations \begin{subequations} \begin{eqnarray}
\left\{ \varphi_{\mathrm{c}}\left(x_{j}\right),\varphi_{\mathrm{d}}^{\dagger}\left(x_{j^{\prime}}\right)\right\} _{+} & = & \delta_{c,d}\delta_{j,j^{\prime}},(\mathrm{c},\mathrm{d}=1,2),\label{a-6-1}\\
\left\{ \varphi_{\mathrm{c}}\left(x_{j}\right),\varphi_{\mathrm{d}}\left(x_{j^{\prime}}\right)\right\} _{+} & = & 0,\label{a-6-2}\end{eqnarray}
corresponds to the slowly varying wavefunctions of the new coordinate
$x_{j}.$ Thus the field operators at $x_{j}$ can be expanded as
\end{subequations} \begin{equation}
\varphi_{\mathrm{c}}\left(x_{j+1}\right)=\varphi_{\mathrm{c}}\left(x_{j}\right)+\left[\frac{\partial\varphi_{\mathrm{c}}\left(x\right)}{\partial x}\right]_{x=x_{j}}2l+\cdots,(\mathrm{c}=1,2).\label{a-7}\end{equation}

The dimerization implies the deformations
$u_{j}^{\mathrm{c}}=\left(-1\right)^{j}\phi_{j}^{\mathrm{c}}$, which
leads the displacement order parameters\begin{eqnarray}
\Delta^{\mathrm{c}}\left(x_{j}\right) & = & -4\alpha\phi^{\mathrm{c}}\left(x_{j}\right)\notag\\
 & = & -2\alpha\left(\phi_{2j}^{\mathrm{c}}+\phi_{2j+1}^{\mathrm{c}}\right).\label{a-8}\end{eqnarray}
As $2N$ is very large, the summation can be substituted with the
integral \begin{equation}
\sum_{j=0}^{2N-1}\rightarrow\int_{-\infty}^{\infty}dx/2l.\label{a-9}\end{equation}
as well as the field operators\begin{equation}
\varphi_{\mathrm{c}}\left(x_{j}\right)\rightarrow\sqrt{2l}\varphi_{\mathrm{c}}\left(x\right).\label{a-10}\end{equation}
Obviously, from Eq. (\ref{a-6-1}) and (\ref{a-6-2}), the above field
operators $\varphi_{\mathrm{c}}\left(x\right)$ satisfy the anti-commutate
relations \begin{subequations} \begin{eqnarray}
\left\{ \varphi_{\mathrm{c}}\left(x\right),\varphi_{\mathrm{d}}^{\dagger}\left(x^{\prime}\right)\right\} _{+} & = & \delta_{\mathrm{c,d}}\delta\left(x-x^{\prime}\right),(\mathrm{c,d=0,1}),\label{a-11-1}\\
\left\{ \varphi_{\mathrm{c}}\left(x\right),\varphi_{\mathrm{d}}\left(x^{\prime}\right)\right\} _{+} & = & 0.\label{a-11-2}\end{eqnarray}

Finally, substituting the Eq. (\ref{a-7})-(\ref{a-10}) into Eq.
(\ref{a-2}) and (\ref{a-3}), the continuum model of the system is
obtained as \end{subequations} \begin{subequations} \begin{eqnarray}
H & = & H_{\mathrm{e}}+H_{\mathrm{p}},\label{a-12-1}\\
H_{\mathrm{p}} & = & \int_{-L}^{L}dx\left\{ \frac{K_{\mathrm{l}}}{8\alpha^{2}l}\Delta^{2}(x)+\frac{M}{32\alpha^{2}l}\overset{\cdot}{\Delta^{2}}(x)\right.+\notag\\
 &  & \left.\frac{K_{\mathrm{t}}}{4^{6}m^{2}\alpha^{4}l}\left[\Delta(x+L)-\Delta(x)\right]^{4}\right\} ,\label{a-12-2}\\
H_{\mathrm{e}} & = & \int_{-L}^{0}dx\Phi(x)\mathscr{H}_{\mathrm{m}}\Phi(x),\label{a-12-3}\end{eqnarray}
where the Hamiltonian density is \end{subequations} \begin{equation}
\mathscr{H}_{\mathrm{m}}=\left[\begin{array}{cc}
iv_{\mathrm{f}}\sigma_{z}\partial_{x}+\Delta(x)\sigma_{x} & G\left(x\right)\\
G\left(x\right) & iv_{\mathrm{f}}\sigma_{z}\partial_{x}+\Delta(x+L)\sigma_{x}\end{array}\right],\label{a-14}\end{equation}
and we have unified the order parameters with $\Delta(x)=\Delta^{\mathrm{a}}\left(x\right)$
and $\Delta(x+L)=\Delta^{\mathrm{b}}\left(x\right),$ with a 4-component
spinor $\Phi(x)=\left[\begin{array}{cccc}
\phi_{1}(x) & \phi_{2}(x) & \phi_{3}(x) & \phi_{4}(x)\end{array}\right]^{\mathrm{T}}$ and modified coupling constant \begin{equation}
G\left(x\right)=G_{0}+\beta\frac{\left(\Delta^{\mathrm{b}}\left(x\right)-\Delta^{\mathrm{a}}\left(x\right)\right)^{2}}{32m\alpha^{2}}.\label{a-15}\end{equation}

\end{document}